\documentclass[aps,prl,twocolumn,superscriptaddress,showpacs]{revtex4}
\usepackage{amssymb}
\usepackage{graphicx}
\usepackage{amsmath}
\usepackage{amsfonts}

\begin{document}

\title{Half Metallic Bilayer Graphene}

 \author{Jie Yuan}
 \affiliation{Department of Physics, and Center of Theoretical and Computational Physics, The University of Hong Kong,
 Hong Kong, China}

\author{Dong-Hui Xu}
\affiliation{Department of Physics, Zhejiang University, Hangzhou,
China}

 \author{Hao Wang}
 \affiliation{Department of Physics, and Center of Theoretical and Computational Physics, The University of Hong Kong,
 Hong Kong, China}
 \affiliation{Department of Physics, South University of Science and Technology of China, Shenzhen, China}

 \author{Yi Zhou}
 \affiliation{Department of Physics, Zhejiang University, Hangzhou,
China}

\author{Jin-Hua Gao}
\email{jinhua@hust.edu.cn} \affiliation{Department of Physics,
Huazhong University of Science and Technology, Wuhan, Hubei, China }
\affiliation{Department of Physics, and Center of Theoretical and
Computational Physics, The University of Hong Kong,
 Hong Kong, China}

 \author{Fu-Chun Zhang}
 \email{fuchun@hku.hk}
 \affiliation{Department of Physics, and Center of Theoretical and Computational Physics, The University of Hong Kong,
 Hong Kong, China}
 \affiliation{Department of Physics, Zhejiang University, Hangzhou,
China}

\begin{abstract}
Charge neutral bilayer graphene has a gapped ground state as
transport experiments demonstrate.  One of the plausible such ground
states is layered antiferromagnetic spin density wave (LAF) state,
where the spins in top and bottom layers have same magnitude with
opposite directions.  We propose that lightly charged bilayer
graphene in an electric field perpendicular to the graphene plane
may be a half metal as a consequence of the inversion and
particle-hole symmetry broken in the LAF state.  We show this
explicitly by using a mean field theory on a 2-layer Hubbard model
for the bilayer graphene.
\end{abstract}

\maketitle Half metals are a class of materials in which electrons
with one spin orientation are metallic and electrons with opposite
spin orientation are insulating~\cite{Groot,Park}. In a half metal,
electrical current can be fully spin polarized. This property is
attractive in spintronics~\cite{Wolf,Murakami,Sinova}. Possibility
of graphene-based half metal has been interesting for its potential
application in electronic devices. Soon after the discovery of
graphene, Son, Cohen and Louie have applied first principles
calculations to propose half metallic phase in a zigzag graphene
nanoribbon with external transverse electric fields~\cite{son}. Due
to the technical difficulties in the devices, the predicted half
metal in graphene ribbon has not been confirmed yet in experiments.
Here we propose that bilayer graphene (BLG) in an electric field
perpendicular to the graphene plane can be a half metal. Our
proposal is based on the recent works to suggest that the ground
state of BLG at half filled (charge neutral case) may be a layered
antiferromagnetic spin density wave (LAF) state with a gap of about
2 meV. In the half filled LAF state, the spins are polarized
oppositely on the top and bottom layers, which break the spin SU(2)
symmetry, but are invariant under a combined transformation of time
reversal ($T$) and inversion ($I$), as we can see in Fig. 1(a). The
LAF state also conserves the combined symmetry of inversion and
particle-hole conjugation ($C$), where the operation of $C$ changes an electron operator to a hole on the same site, which also reverses
the sign of the magnetic moment of a particle (or hole). The electric field breaks
$I$ and $C$.  Their combination $I \otimes C$ is conserved in the neutral graphene, but is broken in the charged graphene, under the electric field.  Therefore, an electric field applied to the charged graphene  opens the
possibility for a half metallic phase with a net magnetization. We
examine the half metallic BLG by applying a mean field theory on a
2-layered Hubbard model. Our prediction may be tested in the BLG
device with double gates\cite{jr}.

We start with a brief summary of the recent works on the
BLG\cite{rmp,castro,jbo,mccann, dft,
ohta,yzhang,mayorov,weiss,jr,yacoby,wbao,freitag,veli,nillson,silva,
hmin,levitov,jung,fzhang,fzhang12,kyang,lemonik,vafek,vafek12,knov,honerkamp,lang,gorbar,ting,varma}.
Theoretically, in the single electron band picture the BLG is a
gapless semiconductor with parabolic valence and conduction bands
touching at the high symmetry points $K$ and $K'$. The state is
unstable in the presence of electron interaction. Experimentally,
there are clear evidences that the BLG at half filled has a gapped
ground state\cite{jr,yacoby,wbao,freitag,veli,weiss}. Velasco et al.
\cite{jr} have applied a perpendicular electric field on a high
quality suspended BLG. The energy gap is found to decrease as the
field increases and to close at a field $~ 15 mV \textrm{nm}^{-1}$.
Two of the most promising states consistent with the gapped ground
states measured in transport experiments are the
LAF~\cite{fzhang,fzhang12,honerkamp,xu,yong} and the quantum spin
Hall states~\cite{kane,raghu,honerkamp,fzhang12,fzhang}. It will be
important and interesting to explore the possible experimental
consequences of these states and determine the true ground state of
the BLG.  In the present Letter, we predict a half metallic phase in
the LAF state of the BLG in the presence of an electric field at
slightly charged graphene. Our result may be used to resolve the
controversial issue of the ground state in BLG, and should be of
importance to the potential application of the graphene in
spintronics.

We consider a BLG system in an applied perpendicular electric field.
The Hamiltonian is given by
\begin{eqnarray}
H=H_0 + H_U + H_p
\end{eqnarray}
where $H_0=H_{intra}+H_{inter}$ is the kinetic energy part, $H_U$ is
the on-site Coulomb interaction, and $H_p$ describes the electric
potential due to the applied electric field. The intra-layer hopping
is given by
\begin{equation}
H_{intra}=-t\sum_{l\langle ij \rangle\sigma}
[a^{\dag}_{l\sigma}(i)b_{l\sigma}(j) + H.c.] + \mu \sum_{li\sigma}
n_{l\sigma}(i)
\end{equation}
Here, $a_{l\sigma}$ ($b_{l\sigma}$) are the electron annihilation
operator on sublattice A (B), and $l=1,2$ for the bottom and top
layers as illustrated in Fig. 1, $\sigma=\uparrow,\downarrow$ and
$i$ ($j$) denote the  spin and site, respectively.  $\langle
\rangle$  sums over all nearest neighbor sites. Here we only
consider the nearest neighbor hopping for simplicity and expect the
small remote hopping will not change the basic physics.  $\mu$ is
the chemical potential. $\mu=0$ corresponds to the half filling, or
the charge neutral point.  At $\mu> 0$, the chemical potential
acroses the conduction band.  We consider an interlayer hopping
between two sites $i$ and $i'$ on-top to each other,
\begin{equation}
H_{inter}=t_{\perp}\sum_{\langle i i'\rangle\sigma}
[b^{\dag}_{1\sigma}(i)a_{2\sigma}(i') + h.c.].
\end{equation}
The Hubbard U term is given by $H_U=U\sum_{li}n_{l\uparrow}(i)
n_{l\downarrow}(i)$. The effect of the external electric field $E_0$
is modeled by an electric potential $V$ between the two layers,
\begin{equation}\label{potential}
H_p=\sum_{li \sigma} V_l n_{l\sigma}(i),
\end{equation}
with $V_l= (-1)^{l}V/2$, and $V$ is related to $E_0$ as below.  Note
that the electron charge on the two layers may be redistributed in
the presence of $E_0$\cite{mccann,dft}. Let $E$ be the electric
field between the two layers, and assume the graphene sheets to be
infinitely large 2-dimensional planes, then we have $V=+ed_0E$, with
$-e$ the electron charge, $d_0$ the interlayer distance. $E$ is
related to $E_0$ by
\begin{equation}\label{field}
E=E_0 - 2\pi e(\rho_2-\rho_1),
\end{equation}
where the electron density in layer $l$ is given by $\rho_l=\langle
\sum_{\sigma,i}n_{l\sigma}(i) \rangle/S$, with $S$ the area of each
layer, and $\langle Q \rangle$ is the average value of operator $Q$.
In the above equation, we have assumed the dielectric constant for
the BLG to be 1 as suggested in the literature\cite{mccann,soff}.

We use a mean field approximation for the Hubbard term, and solve
the Hamiltonian self-consistently,
\begin{equation}\label{Uterm}
H^{\textrm{MF}}_U = U\sum_{li\sigma} \langle n_{l\sigma}(i) \rangle
n_{l \bar{\sigma}}(i)
\end{equation}
There are four atoms in a unit cell and eight mean fields in our
theory, $\langle n^\eta_{l\sigma}\rangle$, with $\eta=A$ or $B$
indicating the sublattice. We use $d_0=0.334$ nm and $2\pi e^2 d_{0}
\approx 3 \times 10^{-11} \textrm{meV}\cdot \textrm{cm}^2$

\begin{figure}
\centering
\includegraphics[width=8cm]{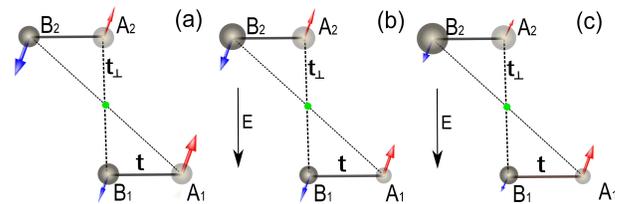}
\caption{(Color online). Schematic illustration of spin (arrows) and
charge (circles) structures of the LAF state in BLG. $A$ and $B$
indicate sublattices, $1$ and $2$ are layer indices. (a) At half
filled and E=0, the state is invariant under both combined
transformations $T\otimes I$ and $I\otimes C$, with $T$, $I$, and $C$ the time reserval, inversion, and particle-hole transformation operators, respectively. (b) At half filled
and finite E, the state is invariant under $I\otimes C$. The charge
structure breaks $I$, but the spin structure remains invariant under
$T\otimes I$. (c) Charged (electron) graphene in
electric field:  spin structure breaks $T\otimes I$ and the state is
ferrimagnetic. }

\end{figure}

At $E_0=0$,  our theory gives a gapped LAF ground state at half
filled, consistent with the experiment~\cite{jr}, and
 with previous theoretical works by using
renormalization group theory\cite{honerkamp,vafek12} and quantum
Monte Carlo method\cite{lang}. First principles calculation
indicates that such LAF state is stable in the presence of nonlocal
Coulomb interaction and remote hopping\cite{yong}.   In Fig. 1(a),
we illustrate the spin and charge structures of the LAF state.  The
charge distribution is uniform and the spins are  anti-parallel to
each other. There are two-fold degeneracy for the spin
configurations, related to the time reversal. The net spin
polarizations in the top and bottom layers have an opposite sign,
and their sum gives a null magnetization. In Fig. 2(a), we show the
LAF energy gap as a function of $U$.  For  $U \approx 6.64$eV, we
have a gap of  $\epsilon_g \approx 2$ meV, approximately the gap in
the BLG measured in the experiment. Note that the estimated value of
$U$ here for the BLG is close to the value of $U \approx 6.2$eV
estimated to fit the experimentally observed energy gap in the ABC
stacking trilayer graphene\cite{xu}. Therefore, the model seems to
gives a comprehensive picture about the energy gap of the few layer
graphene systems\cite{xu,tri}.

\begin{figure}
\includegraphics[width=8cm]{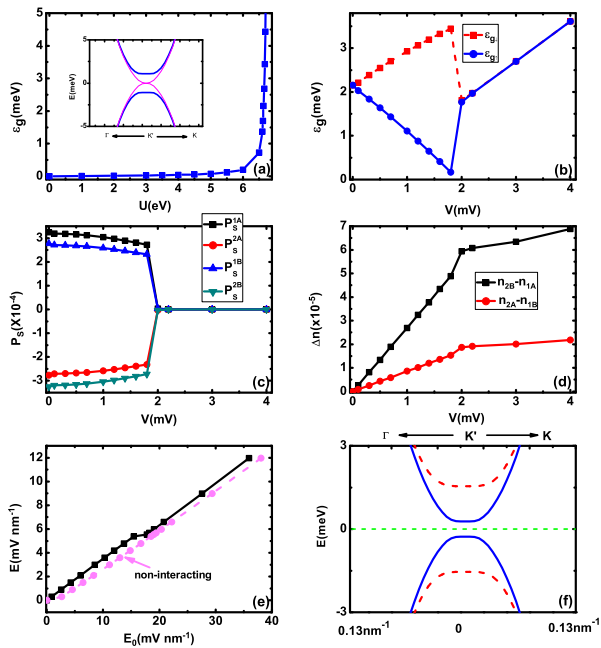}
\caption{(Color online). Bilayer graphene at half filled.  (a)
Energy gap of $H$ in Eqn. (1) as a function of Hubbard U. Shown in
inset are low energy bands (blue curves).  The parameters are
$U=6.64$eV, $t=3.16$eV, and $t_\perp=0.381$ eV.  The bands for $U=0$
are plotted for comparison (red curves). (b) Spin-resolved energy
gap as function of V. (c) spin polarization on four distinct lattice
sites. (d) Charge transfer as functions of interlayer electric
potential $V$.  In (c) and (d), superindices 1(2) and A (B) are for
layer indices and sublattices, respectively. (e) Electric field $E$
as a function of applied electric field $E_0$ (solid black line),
also shown is $E$ v.s. $E_0$ for non-interacting case  $U=0$ (dashed
pink line). (f). Energy bands at $V=1.4$meV. }
\end{figure}

We proceed to discuss the effect of the electric field to the half
filled BLG. The real space spin and charge distributions are
illustrated in Fig. 1(b). At half filled,  an electric field breaks
inversion symmetry $I$, and also the particle-hole conjugation
symmetry $C$, but conserves $I\otimes C$. This can be seen from the
Hamiltonian for $H_p$.  The half filled LAF state in the electric
field is invariant under $I \otimes C$. Namely, the charge is
invariant under $I\otimes C$, while the spin remains invariant under
both $T\otimes I$ and $I\otimes C$. The system has no net
magnetization.  In Fig. 2(b), we plot the calculated spin dependent
energy gap  as a function of $V$.  As $V$ increases, the spin-down
excitation gap increases and the spin-up excitation gap becomes
narrowed.  At a critical value $V=V_c \approx 1.8 \textrm{mV}$ ,
there is a first order transition from the LAF state to the spin
symmetric normal state. There is a sudden change of the gap at
$V_c$.   At $V> V_c$ the gap increases monotonously with $V$. In
Fig. 2(c,d), we show the spin polarization and the charge transfer
at the four distinct lattice sites. Our calculations confirm the symmetry analysis we have made
about the spin and charge structure. There is a charge transfer
from sublattice A in layer 1 to sublattice B in layer 2, and a less
amount of charge transfer from sublattice B in layer 1 to sublattice
A in layer 2 (inter-layer nearest neighboring sites).  In Fig. 2(e),
we show the screening effect and plot $E$ as a function $E_0$,
according to Eq. (4). We remark that $V_c\approx 1.8 mV$ corresponds
to an applied electric field $E_0 \approx 15 \textrm{mVnm}^{-1}$.
This is in good agreement with the external electric field at the
phase transition point estimated in the transport
measurement~\cite{jr}.

In Fig. 2(f) we show the spin-dependent bands near the Fermi level
for $V=1.4$ meV.  Both the lower energy valence and conduction bands
are for spin-up electrons, and the spin-down electron bands are
further away from the Fermi level. Because of its semiconducting
nature, all the valence bands are occupied and all the conduction
bands are empty at zero temperature, and the ground state is spin
non-polarized.   The spin-polarized low energy excitations can be
detected in spin-polarized transport experiment, and the gap
measured by Velasco et al.~\cite{jr} in the presence of external
electric field should be spin polarized, in agreement with previous
work ~\cite{fzhang12}. The spin polarization of the excitation gap
can be detected in transport measurement with ferromagnetic source
and drain electrodes.

\begin{figure}
\includegraphics[width=8cm]{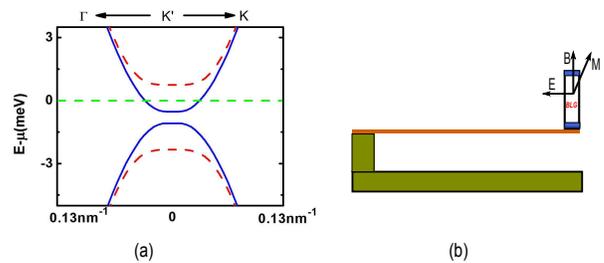}
\caption{(Color online). (a) Energy bands of the BLG in an
interlayer electric potential $V=1.4$ mV at a charged graphene with
electron density  $\delta n \approx 2 \times 10^{10}
\textrm{cm}^{-2}$. Solid lines are for spin-up and dashed ones for
spin-down. Model parameters are the same as in Fig. 2.  (b):
schematics of probing magnetization by using torque magnetometry
experiment.
 }
\end{figure}

The most interesting case is at $\mu >0$ but small, where we have a
metallic state in the background of LAF. The metallic state in a
perpendicular electric field can be a half metal.  The spin and
charge structures of this case is illustrated in Fig. 1(c) and there
is a net spin-up in the BLG. Away from half filled, there is no
particle-hole symmetry, and the system in the electric field breaks
$I$, $C$, and $I\otimes C$, and the spin structure of the LAF state
breaks $T \otimes I$ symmetry and gives a half metal with a net
magnetization. The case for $\mu <0$ can be obtained by
particle-hole transformation and will not be discussed further.
Experimentally, the shift on the chemical potential can be realized
by tuning the gate voltage, which may be controlled independently
together with the tuning of the electric field in double gated BLG
device.  In Fig. 3(a), we plot the low energy bands for the BLG at
the lightly charge density $\delta n \approx 2 \times 10^{10}$
cm$^{-2}$.  The other  parameters are the same as in Fig. 2(f). This
electron density is equivalent to $2.5 \times 10^{-6}$ electron per
carbon site on the BLG.  As we can see from the figure, the spin-up
conduction band is partially filled but the spin-down conduction
band is completely empty. The system is a half metal with a full
spin polarization in its carriers.  The electric field induces time
reversal symmetry broken in this case. The surface magnetization per
area is $M = \delta n \times \mu_G = 2 \times 10^{-4} g_L\mu_B/2$
per nm$^2$, with $\mu_G$ the magnetic moment of graphene atom, 
and
$\mu_B$ the Bohr magneton and $g_L$ the Lande g-factor for the
graphene, which is about 2-2.5.  The magnetization is tiny, but possibly detectable by
using torque magnetometry experiment, as schematically illustrated
in Fig. 3(b).~\cite{Lilu}

\begin{figure}
\includegraphics[width=8cm]{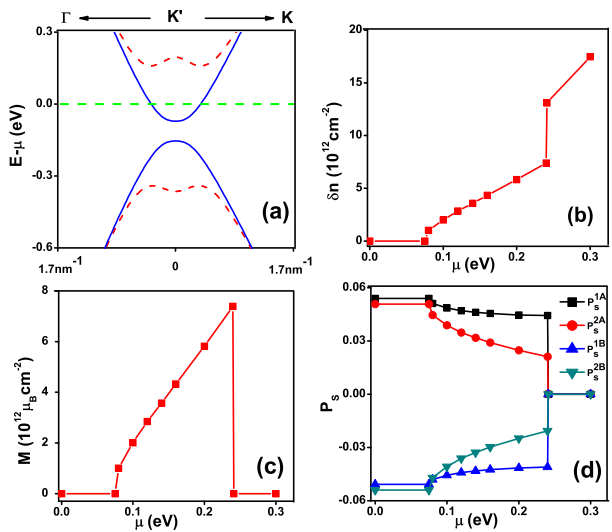}
\caption{(Color online). BLG in an interlayer electric potential
$V=0.25$ eV for model parameters $U=7.2$ eV, $t=3.16$ eV, and
$t_{\perp} =0.5$ eV. (a). Energy bands at an electron doping $\delta
n \approx 2 \times 10^{12} \textrm{cm}^{-2}$. Blue solid lines are
for spin-up and red dashed ones for spin-down. (b). Electron density
$\delta n$ vs chemical potential.  The jump in $\delta n$ reflects
first order phase transition from LAF to normal states, and the
system is phase separated in the "jumped" density region.  (c).
Magnetization as function of chemical potential ($g_L = 2$; and (d). Site spin
polarizations as functions of chemical potential.
 }
\end{figure}

We have studied $\mu$ or $\delta n$ dependence of the magnetization
and the spin polarization of the BLG away from the half filled.  As
$\mu$ increases, there is a jump in $\delta n$ at a critical $\mu_c$
to separate the LAF and normal states.  To illustrate this,  we
consider model parameters given in Fig. 4, which gives a larger LAF
gap at half filled, hence a more pronounced half metallic state when
it is doped. The jump in $\delta n$ as a function of $\mu$ can be
seen clearly in Fig. 4(b).  The magnetization and spin polarization
are plotted in Fig. 4(c) and (d), respectively. At $\delta n=0$, the
spin polarization $P_S^{1A} = -P_S^{2B}$, and $P_S^{2A} =
-P_S^{1B}$.  At $\mu >0$,  we find $P_S^{1A} > |P_S^{2B}|$, and
$P_S^{2A} >|P_S^{1B}|$, which results in a net magnetization and is
consistent with the band picture in Fig. 4(a).

In conclusion, we have studied the LAF state in the BLG by using a
mean field theory.  At half filled, a weak perpendicular electric
field leads to spin polarized low energy excitations.  In slightly
charged graphene,  the electric field can induce a half metallic
phase with conduction electron or valence hole completely spin
polarized, and the system is ferrimagnetic.   The required strength
of the electric field and the range of the electron density are
within the reach of the present experiments in double gated BLG.
Detecting the nonzero magnetization should be a feasible way to
identify the half metallic phase in the BLG. Such an experiment will
also provide direct evidence for the LAF state.  Our results on the
half metallic phase in the BLG are robust, and are the consequence
of the symmetry in the LAF state in connection with the inversion
symmetry broken induced by an electric field and particle-hole
symmetry broken away from half filled.   From the band theory point
of view, the low energy spin-up and spin-down bands of the LAF state
in the half filled BLG are split by the electric field, which leads
to a half metal once the carriers are generated in gating.  Finally,
we argue that such half metallic phase should be found in the
N-layer graphene with rhombohedral stacking ordering, where the LAF
state has been proposed to be the ground state as we reported
recently~\cite{xu}.   The LAF gap due to the electron interaction is
larger for larger N, and saturates at value of about $20$ meV around
$N=9$.   The half metallic phase in multilayer graphene  may be more
robust than in the BLG, but perhaps more challenging to realized in
experiments.  Theoretical work along this line is on-going.

We thank W. Q. Chen for many useful discussions and
L Li for a helpful discussion on the possibility to use torque
magnetometry to measure magnetization in layered systems. We
acknowledge partial financial support from HKSAR RGC Grant No. HKU
701010 and CRF Grant No. HKU 707010.  J.H.G is supported by the
National Natural Science Foundation of China (Project No.11274129).
D.H.X and Y.Z. supported by National Basic Research Program of China
(973 Program, Grant No.2011CBA00103), the NSFC (Grant No.11074218)
and the Fundamental Research Funds for the Central Universities in
China.


\begin{thebibliography}{}
\bibitem{Groot} R. A. de Groot, F. M. Mueller, P. G. van Engen and K. H. Buschow,  Phys. Rev. Lett. 50, 2024 (1983).

\bibitem{Park} J. H. Park et al.  Nature 392, 794 (1998).

\bibitem{Wolf} S. A. Wolf et al.  Science 294, 1488 (2001).

\bibitem{Murakami} S. Murakami, N. Nagaosa, and S. C. Zhang, Science 301, 1348 (2003).

\bibitem{Sinova} J. Sinova et al. Phys. Rev. Lett. 92, 126603 (2004).

\bibitem{son} Y. W. Son, M. L. Cohen, and S. G. Louie, Nature  444, 347 (2006).

\bibitem{jr}
J. Velasco Jr., L. Jing, W. Bao, Y. Lee, P. Kratz, V. Aji, M.
Bockrath, C. N. Lau, C. Varma, R. Stillwell, D. Smirnov, F. Zhang,
J. Jung, and A. H. MacDonald, Nature Nanotech. 7, 156 (2012).

\bibitem{rmp}
A.H.C. Neto, F. Guinea, N.M.R. Peres, K.S. Novoselov, and A.K. Geim,
Rev. Mod. Phys. 81, 109 (2009).

\bibitem{ohta}
T. Ohta, A. Bostwick, T. Seyller, K. Horn, E. Rotenberg, Science
313, 951 (2006).
\bibitem{jbo}
J. B. Oostinga, H. B. Heersche, X. Liu, A. F. Morpurgo, and L. M. K.
Vanderspen, Nature Mater. 7, 151 (2008).

\bibitem{castro}
E. V. Castro, K. S. Novoselov, S. V. Morozov, N. M. R. Peres, J. M.
B. Lopes dos Santos, J. Nilsson, F. Guinea, A. K. Geim, and A. H.
Castro Neto, Phys. Rev. Lett. 99, 216802 (2007).

\bibitem{yzhang}
Y. Zhang, T.-T. Tang, C. Girit, Z. Hao, M. C. Martin, A. Zettl, M.
F. Crommie, Y. R. Shen, and F.Wang, Nature 459, 820 (2009).

\bibitem{mccann}
E. McCann, Phys. Rev. B 74, 161403(R) (2006).

\bibitem{dft}
H. Min, Bhagawan Sahu, Sanjay K. Banerjee, and A. H. MacDonald,
Phys. Rev. B 75, 155115 (2007).

\bibitem{mayorov}
A. S. Mayorov, D. C. Elias, M. Mucha-Kruczynski, R. V. Gorbachev, T.
Tudorovskiy, A. Zhukov, S. V. Morozov, M. I. Katsnelson, V. I. Fal
¡äko, A. K. Geim, and K. S. Novoselov, Science 333, 860 (2011).

\bibitem{weiss}
F. Freitag, M. Weiss,. R. Maurand, J. Trbovic, and C.
Sch\"{o}nenberger, arXiv:1212.5918.

\bibitem{yacoby}
R. T. Weitz, M. T. Allen, B. E. Feldman, J. Martin, and A. Yacoby,
Science 330, 812 (2010).

\bibitem{freitag}
F. Freitag, J. Trbociv, M. Weiss, and C. Sch¡§nenberger, Phys. Rev.
Lett. 108, 076602 (2012).

\bibitem{veli}\label{gap}
 A. Veligura, H. J. van Elferen, N. Tombros, J. C. Maan, U.
Zeitler, and B. J. van Wees, Phys. Rev. B 85, 155412 (2012).

\bibitem{wbao}
W. Bao, J. Velasco Jr., L. Jing, F. Zhang, B. Standley, D. Smirnov,
M. Bockrath A. H. MacDonald, and C. N. Lau, Proc. Natl. Acad. Sci.
USA 109, 10802 (2012).

\bibitem{nillson}
J. Nilsson, A. H. Castro Neto, N. M. R. Peres, and F. Guinea, Phys.
Rev. B 73, 214418 (2006).


\bibitem{silva}
E. V. Castro, N. M. R. Peres, T. Stauber, and N. A. P. Silva, Phys.
Rev. Lett. 100, 186803 (2008).

\bibitem{hmin}
H. Min, G. Borghi, M. Polini, and A. H. MacDonald, Phys. Rev. B 77,
041407(R) (2008).

\bibitem{levitov}
R. Nandkishore and L. Levitov, Phys. Rev. Lett. 104, 156803 (2010);
Phys. Rev. B 82, 115124 (2010).

\bibitem{jung}
J. Jung, F. Zhang, and A. H. MacDonald, Phys. Rev. B 83, 115408
(2011).

\bibitem{fzhang}
F. Zhang, J. Jung, G. A. Fiete, Q. Niu, and A. H. Mac- Donald, Phys.
Rev. Lett. 106, 156801 (2011);

\bibitem{fzhang12}
F. Zhang and A. H. MacDonald, Phys. Rev. Lett. 108, 186804 (2012).

\bibitem{kyang}
O. Vafek and K. Yang, Phys. Rev. B, 81, 041401 (2010).

\bibitem{lemonik}
Y. Lemonik, I. L. Aleiner, C. Toke, and V. I. Fal¡¯ko, Phys. Rev. B
82, 201408 (2010); Y. Lemonik, I. L. Aleiner, and V. I. Fal¡¯ko,
Phys. Rev. B 85, 245451 (2012).

\bibitem{vafek}
O. Vafek, Phys. Rev. B, 82, 205106 (2010).

\bibitem{knov}
M. Kharitonov, Phys. Rev. Lett. 109, 046803 (2012); arXiv:1109.1153.


\bibitem{vafek12}
R. E. Throckmorton, O. Vafek, Phys. Rev. B 86, 115447 (2012).




\bibitem{honerkamp}
M. M. Scherer, S. Uebelacker, and C. Honerkamp, Phys. Rev. B 85,
235408 (2012).
\bibitem{lang}
T. C. Lang, et al., Phys. Rev. Lett. 109, 126402 (2012).

\bibitem{gorbar}
E. V. Gorbar, V. P. Gusynin, V. A. Miransky, and I. A. Shovkovy,
Phys. Rev. B 85, 235460 (2012).

\bibitem{ting}
Y. Z. Yan, C. S. Ting,   arXiv:1210.0166.

\bibitem{varma}
Lijun Zhu, Vivek Aji, Chandra M. Varma, Phys. Rev. B 87, 035427
(2013)

\bibitem{kane}
C. L. Kane and E. J. Mele, Phys Rev. Lett. 95, 226801 (2005).

\bibitem{raghu}
S. Raghu, X. L. Qi, C. Honerkamp, and S. C. Zhang, Phys. Rev. Lett.
100, 156401 (2008).

\bibitem{xu}
Dong-Hui Xu, Jie Yuan, Zi-Jian Yao, Yi Zhou, Jin-Hua Gao, and
Fu-Chun Zhang, Phys. Rev. B 86, 201404 (2012).



\bibitem{yong}
Yong Wang, Hao Wang, Jin-hua Gao, Fu-Chun Zhang, arXiv:1301.0052.

\bibitem{soff}
F. L¨¦onard and J. Tersoff, Appl. Phys. Lett. 81, 4835 (2002).

\bibitem{tri}
M. M. Scherer, S. Uebelacker, D. D. Scherer, and C. Honerkamp, Phys.
Rev. B 86, 155415 (2012).

\bibitem{Lilu}  Lu Li,  C. Richter,  J. Mannhart, and  R. C. Ashoori, Nature Physics 7, 762-766 doi:10.1038/nphys2080.



\end{thebibliography}
\end{document}